\documentclass[manuscript]{JASA}

\usepackage{hyperref}

\usepackage{academicons}
\usepackage{setspace} 

\usepackage{multirow}%
\usepackage{amsmath,amssymb,amsfonts}%
\usepackage{booktabs}%
\usepackage{algorithmicx}%
\usepackage{algpseudocode}%

\usepackage{orcidlink}

\usepackage{placeins}

\begin{document}
\nolinenumbers
\renewcommand{\linenumberfont}{\relax}
\renewcommand{\thelinenumber}{}

\title[]{Classifying bioacoustic data without individual call annotations using temporal convolutional networks and feature extractors}
\author{Laia Garrobé Fonollosa $^{\orcidlink{0009-0002-9383-4815}}$} 
\email{lgf3@st-andrews.ac.uk}

\affiliation{Sea Mammal Research Unit, School of Biology, University of St. Andrews, KY16 9TH, St. Andrews, Scotland}

\author{Douglas Gillespie$^{\orcidlink{0000-0001-9628-157X}}$}

\affiliation{Sea Mammal Research Unit, School of Biology, University of St. Andrews, KY16 9TH, St. Andrews, Scotland}
 
 \author{Lina Stankovic$^{\orcidlink{0000-0002-8112-1976}}$		}	

\affiliation{Electrical and Electronic Engineering, University of Strathclyde, Montrose Street, Glasgow, G1 1XJ,  Scotland}
 
\author{Vladimir Stankovic $^{\orcidlink{0000-0002-1075-2420}}$		}	

\affiliation{Electrical and Electronic Engineering, University of Strathclyde, George Street, Glasgow, G1 1XW,  Scotland}

\author{Luke Rendell $^{\orcidlink{0000-0002-1121-9142}}$}

\affiliation{Sea Mammal Research Unit, School of Biology, University of St. Andrews, KY16 9TH, St. Andrews, Scotland}


\date{\today}

\begin{abstract}
Bioacoustic data from Passive Acoustic Monitoring (PAM) generates large datasets where obtaining detailed auditing and labelling is often impractical, resulting in weak annotations (e.g., presence/absence of species over several minutes of recording). In order to effectively capture the complex temporal patterns and key features of long audio segments, we propose a framework comprising dataset standardisation, feature extraction, and classification via Temporal Convolutional Networks (TCN). This approach eliminates the necessity for setting heuristic decision rules or creating time-consuming strong labels. To demonstrate the effectiveness of our approach, we use sperm whale (\textit{Physeter macrocephalus}) click trains in 4-minute recordings as a case study, from a dataset comprising diverse sources and deployment conditions to maximise generalisability. Our TCN classifiers achieve recall rates exceeding 0.83 at a 0.13 false positive rate, comparable to agreement rates between expert annotators.  We compare two methods of feature extraction, Variational AutoEncoders (VAEs) and traditional handpicking of features, and found them to yield similar performance results, with the VAE-based classifiers seeing a more stable performance across datasets and recording conditions. 
These results offer a way forward in leveraging numerous existing annotated bioacoustic datasets to train automatic classification models, effectively overcoming previous limitations associated with weak labels.
\end{abstract}


\maketitle


\section{\label{sec:1} Introduction}

Passive Acoustic Monitoring (PAM) uses acoustic recorders to survey and monitor wildlife and environments. In the case of marine animals, PAM offers the advantage of enabling continuous data collection, even under adverse weather conditions or when visual surveys are unfeasible. As our capacity for collecting large quantities of acoustic data continues to grow, so does the demand for automated detection systems to effectively process and analyse them. While PAM has traditionally relied on many different computational techniques for data analysis \citep{mellinger_recognizing_2000, gradisek_predicting_2017, fagerlund_bird_2007}, Deep Learning (DL) methods have gained popularity for tasks that require detection of animal sounds from large acoustic datasets \citep{stowell_computational_2022}. 
 
Underwater PAM datasets pose a unique set of challenges when it comes to training DL methods, arising from the lack of reliable ground-truthing, the substantial cost associated with data collection and auditing, and the scarcity of target events. This results in considerably smaller annotated datasets with few instances of each signal of interest, both orders of magnitude below the dimensions of typical benchmark datasets \citep{von_benda-beckmann_northern_2022, xeno-canto_xeno-canto_2024, krizhevsky_imagenet_2017}. Moreover, bioacoustic baseline datasets are often derived from limited surveys using single equipment types and narrow spatiotemporal ranges \citep{dclde_dataset_2015, dclde_dataset_2018, dclde_dataset_2024}. While these datasets are very valuable to establish a solid baseline for DL biacoustic studies, their reduced scope limits the generalisability of models trained \citep{napoli_unsupervised_2023}. Finally, much of underwater PAM data are annotated at a coarse temporal resolution relative to the individual vocalisations, a form of weak labelling. While weak labelling is a term that carries conflicting definitions across fields, here we adopt the broader interpretation of annotation formats that do not provide precise per-call boundaries. These include presence/absence labels specifying whether vocalisations occur within fixed-duration intervals (often a few minutes), as well as acoustic event boundaries (e.g., species acoustic encounters). In both cases, individual calls are not explicitly delineated, and there may be periods of silence within the annotation period. These practices balance cost-effectiveness with field practicality, while also reducing variability between annotators \citep{DBLP:conf/dcase/NapoliWB22}. However, weak labelling approaches considerably increase computational demands by generating datasets that are larger and more complex than those produced by strong labels. In turn, weakly labelled datasets require  larger and more complex models to process compared to finely annotated data, often rendering it practically infeasible.

Weak labelling is particularly useful for analysing sperm whale (\textit{Physeter macrocephalus}) vocalisations. Sperm whales emit frequent and easily detectable clicks for the majority of their dive time \citep{madsen_male_2002, watwood_deep-diving_2006}. Maximum source levels of 240dB re 1 µPa have been recorded \citep{mohl_monopulsed_2003}, with click lengths of 1ms and average inter-click intervals of around 0.5 seconds for regular vocalisations \citep{whitehead_click_1990}. The regularity of these clicks makes PAM particularly suitable to study sperm whales, and because they are widely distributed, there are dozens of underwater datasets featuring their vocalisations. Sperm whale's sound output is directional and varies with the animal's aspect \citep{madsen_male_2002}, and what is received is also affected by sound propagation.  Therefore, the acoustic and spectral properties of individual sperm whale clicks contain limited information to distinguish them from other impulsive sounds, particularly at low signal-to-noise ratios (SNR), but their regularity in terms of amplitude and inter-click-interval provides crucial contextual information. This makes weak labelling a practical choice for annotating sperm whale acoustic data, as it enables the annotator to leverage both local information on possible individual click characteristics and broader temporal patterns when making detection decisions. An effective detection system should therefore integrate both levels of spatiotemporal information to reliably identify click trains.

A suite of algorithms have been developed to detect sperm whale echolocation clicks, including approaches based on the time-frequency analysis of the clicks \citep{miller_seasonal_2018, morrissey_passive_2006}, energy comparisons \citep{klinck_energy_2011}, or the use of the Teager-Kaiser Energy Operator \citep{kandia_detection_2006}. In the field of DL, research has used convolutional neural networks to distinguish between spectrograms containing or not containing  clicks from odontocetes \citep{luo_convolutional_2019} or, more specifically, sperm whales \citep{bermant_deep_2019}. However, such studies focus on detections of individual clicks, rather than incorporating multiple clicks into the decision process, likely ignoring important information encoded in the regularity of click vocalisations. Recent studies have attempted to detect sperm whale click trains through heuristic methods, including \citet{macaulay_passive_2020} for detecting porpoise click trains, later applied to sperm whales by \citet{webber_streamlining_2022}. Another approach used deep learning models to classify long recordings (4 minutes) based on the patterns of click detections across consecutive 5-second windows \citep{garrobefonollosaComparingNeuralNetworks2024}. While both approaches were successful in their respective tasks, the method proposed by \citet{macaulay_passive_2020} requires extensive manual fine-tuning to suit specific datasets, and the DL approach of \citet{garrobefonollosaComparingNeuralNetworks2024} is limited by the requirement for more finely labelled data (in this case, annotations at a 5-second level), which is not always available.

This study proposes a new methodology for detecting sperm whale click trains that addresses two of the most prevalent challenges of bioacoustics data and training DL models for acoustic detection. Firstly, we tackle the issue of small dataset sizes and high variability in environmental and anthropogenic noise and acoustic data across different sources and geographical regions by curating a dataset from multiple sources collected under various deployment conditions. Secondly, we address the problem of weak labelling by proposing a workflow that leverages existing weakly annotated data to train robust DL classification models. This comprises unsupervised feature extraction, for which we compare unsupervised variational autoencoders (VAEs) to expert-led handpicked acoustic features.  This step avoids the need for setting manual thresholds (e.g. amplitude or energy cut-offs) or providing strong labels. The extracted features are then classified via a Temporal Convolutional Network (TCN), a DL architecture tailored for sequence processing \citep{bai_empirical_2018}, which has been successfully employed for acoustic data classification in recent years \citep{bai_empirical_2018, davies_temporal_2019, xie_frog_2022} due to its ability to capture long-term dependencies and temporal context. To the best of our knowledge, this paper is the first to propose the VAE-TCN framework in the context of bioacoustic data detection and classification.

Traditional bioacoustic detection often relies on hand-made parameters such as peak frequencies, spectral shape, and inter-click intervals \citep{usman_review_2020, towsey_use_2014, miller_seasonal_2018}. While recent methods using energy-temporal features \citep{li_classification_2024} or acoustic indices \citep{frasier_machine_2021} show promise, they often fail in new soundscapes \citep{sethi_limits_2023}. Autoencoders \citep{bianco_machine_2019} and VAEs have been shown to surpass expert-selected features for tasks such as species discrimination \citep{goffinet_low-dimensional_2021, rowe_acoustic_2021} and call clustering \citep{reeves_ozanich_unsupervised_2020}, but direct comparisons for complex tasks remain lacking. We chose VAEs over deterministic autoencoders because they encode input data into a probabilistic latent space, which enhances robsustness to variation and noise in the input data, and allows for smooth interpolations in between data points in this continuous latent space \citep{rezende_stochastic_2014}. We assessed the ability of VAEs to learn meaningful patterns and extract information from waveforms and spectrogram representations without prior knowledge of the target sounds. To evaluate the learning efficacy of VAEs, we compared them to traditional, expert-led methods \citep{cox_development_2011} to select acoustic parameters, including root-mean-square calculations, energy level sums, and spectral properties.

In this study, we address key challenges in weakly labelled bioacoustic detection through five main objectives:
\begin{enumerate}
     \item Curate  PAM recordings from various sources and deployment conditions into a comprehensive dataset that includes different types of background noise and other vocalising species to promote generalisation of the solution. 
     Recordings from different studies are re-sampled and preprocessed, with labels standardised and erroneous annotations removed. This dataset is weakly labelled with annotations for periods of multiple minutes, and contains balanced instances of the signal of interest (i.e., sperm whale vocalisations) and anthropogenic and environmental noise.

    \item Develop a VAE-based feature extraction approach to learn acoustic features from recordings without requiring additional, time-consuming and thus costly, labelling, providing an alternative to handpicked acoustic features. 
    
    \item Train a TCN classifier on the extracted features to classify 4-minute long recordings as containing/not containing sperm whale click trains.
    
    \item Evaluate the efficacy of the extracted features in terms of classification accuracy against conventionally employed handpicked acoustic features. 
    
    \item Assess the importance of annotation length on model performance to determine whether a greater emphasis on temporal context, or higher temporal resolution in extracted features, influences performance.
    
\end{enumerate}

\section{\label{sec:2} Methods}

\begin{figure}[!th]
    \centering
    \includegraphics[width=.5\textwidth]{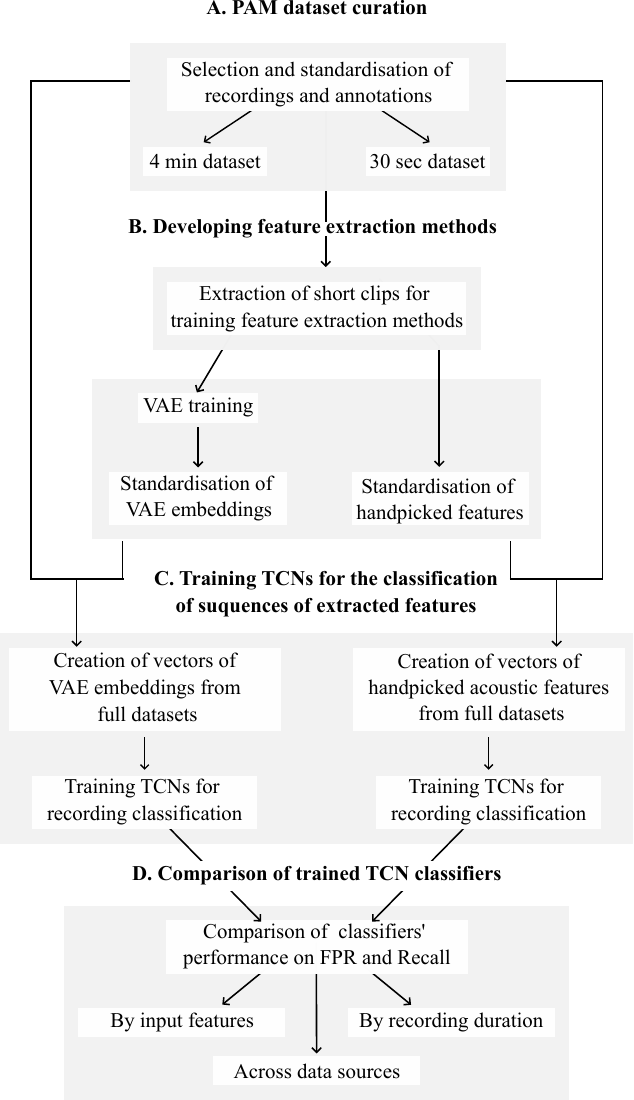}
    \caption{Overview of the methodology for training models on weakly labelled data (i.e., recordings without per-call annotations) to classify recordings based on the presence/absence of sperm whale (\textit{Physeter macrocephalus}) clicks. The four steps include (A) dataset curation and standardisation, (B) developing methods for fine-grained feature extraction using variational autoencoders (VAEs) and handpicked features, (C) training temporal convolutional networks (TCN) for recording-level classification, and (D) comparative evaluation of models using different feature sets. }
    \label{fig:workflow}
\end{figure}

Our proposed framework follows a systematic workflow comprising four stages, as shown in Figure ~\ref{fig:workflow}. Each of the four stages is described next.  

\subsection{PAM dataset curation}
To enhance the robustness and generalisability of the trained models, a comprehensive dataset was compiled from data collected by multiple studies using various recorders and protocols in different types of environment, covering different geographical and temporal scales (Table \ref{tab:sources}). By collating data from various sources, we aimed to ensure robustness to the variabilities in the PAM datasets across various factors that would otherwise impact the classifier’s generalisation ability. These factors included: 
\begin{enumerate}
    \item   Variation in the rhythmic patterns of sperm whale vocalisations, influenced by factors such as the number of individuals and behavioural state.  
    \item Diverse recording conditions and deployment parameters, such as the recorder's depth, proximity to the surface or seafloor, or frequency response of the recording systems.
    \item The presence of anthropogenic and natural ambient sounds, as well as other vocalising species, and signal-to-noise ratio of the recordings.
    \item Annotation variability from individual-level click labelling to 4-minute click train labelling to semi-automatic labelling.
\end{enumerate}

\singlespacing

\begin{table}[]
\footnotesize
    \centering
    \caption{Summary of data sources contributing to the sperm whale dataset, detailing geographic location, temporal scale, recording setup, and labelling approach for source.}
    \begin{tabular}{p{.1\textwidth}p{.18\textwidth}
    p{.08\textwidth}p{.3\textwidth}p{.3\textwidth}}
    \hline
         Dataset name & Location& Time period & Recording setup &  Labelling process  \\
         \hline \hline
            AS & Atlantic, Southern, Arctic and Indian Oceans & 2019 - 2021 & Towed hydrophone sampling at 500kHz. & Individual clicks labelled using PAMGuard automatic detector \citep{gillespie_pamguard}, with all detections reviewed by human analyst. \\     \hline     
            
            BAL\_1 &  Balearic Sea  & 2015 - 2018  & \multirow{2}{.3\textwidth}{Ecological Acoustic Recorders (Oceanwide Science) sampling at 64 kHz for 4 min every 30. Depth  $\sim$ 300m. } & \multirow{4}{.3\textwidth}{4 min recordings labelled as either containing sperm whale click trains, possible single vocalisations, or no sperm whale sounds.}\\
            \cmidrule(lr){1-1} 
            \cmidrule(lr){3-3}
            BAL\_2 &  & 2015 - 2017 &  & \\
            \cmidrule(lr){1-1} \cmidrule(lr){3-4}
            BAL\_3 &  & 2018 - 2019  & SoundTraps  (Ocean Instruments NZ) sampling at  96 kHz for 4 min. every 30. Depth $\sim$ 300m. &   \\
            \hline

            CAL &  Southern California Bight & 2009 - 2013 & High-frequency acoustic packages \citep{wiggins_high-frequency_2007} sampling at 100kHz - 160 kHz. Depth $\sim$ 65m-1300m & Start and end time of odontocete encounters annotated. \\ \hline
            
            CS & Caribbean Sea & 2019 - 2020 & Custom built towed hydrophone sampling continuously at 96 kHz. Depth $\sim$ 10m.  & 4 min every 30 audited in the field. Cetaceans and anthropogenic sounds annotated for the whole period. \\      \hline

            ICE & Iceland & 2021 - 2022 & Stereo deep water recorders (Loggerhead Instruments LS2X) at depth $\sim$ 2300m sampling at 96kHz for 5 min every 15. &   Files that triggered a peak-to-peak SPL detector were manually audited and annotated. All clicks in each file were assigned the same label. \\ \hline
            
            MED & Mediterranean Sea & 2004 - 2005 & Towed array sampling at 48 kHz. Depth $\sim$ 10m. &  Individual clicks manually annotated using Rainbowclick software \citep{ifaw_rainbowclick_2011}. \\ \hline

    \end{tabular}
    \label{tab:sources}
\end{table}

\doublespacing

The main dataset used in this study comprises data from six studies conducted in  Atlantic, Southern, Arctic and Indian oceans (AS dataset ; \citealt{webber_streamlining_2022}), the Balearic Sea (BAL datasets; \citealt{garrobefonollosaComparingNeuralNetworks2024}), the California Bight (CAL dataset; \citealt{dclde_dataset_2015}), the Caribbean Sea (CS dataset; \citealt{vachonOceanNomadsIsland2022}), the Icelandic off-shore waters (ICE dataset; \citealt{wensveen2022}) and the Mediterranean Sea (MED dataset; \citealt{lewis_abundance_2018}). Data from the BAL source were divided into its 3 deployments (BAL\_1, BAL\_2, BAL\_3) given that there was enough data available and previous studies showed high variability in acoustic conditions between deployments \citealt{garrobe_seasonal_2021}.

Data collection varied across sources. The AS, CS and MED sources were collected using towed hydrophones, while the BAL, CAL, CS and ICE sources used moored autonomous recorders. The detail and quality of the annotations also varied among datasets, including individual-level click labelling (MED dataset), labelling at a 4-minute level (BAL and CS datasets), or semi-automatic labelling using automated acoustic algorithms to do an initial processing of the data (AS and ICE datasets). Across all sources, labelling was conducted using both listening and a variety of visual displays (e.g., spectrograms). 

All data were collected and annotated for other studies. We did not re-evaluate the original labels or conduct further annotation efforts, as this was not feasible given the scale and diversity of the datasets. Instead we accepted the labels as provided, acknowledging some level of label inaccuracy. However, this diversity of annotation practices also serves to train the model to generalise across varying  labelling conditions. Table \ref{tab:sources} summarises the key characteristics of each source and labelling process. Detailed descriptions of site locations, other species presence, and data collection protocols, can be found in the original publications \cite{webber_streamlining_2022, garrobefonollosaComparingNeuralNetworks2024, dclde_dataset_2015, vachonOceanNomadsIsland2022, wensveen2022, lewis_abundance_2018}.

Given the diverse recording conditions and labelling formats across data from different sources, a necessary first step was standardising both recordings and labels. A uniform labelling format was established, which entailed dividing all recordings into 4-minute segments that were then labelled either as either containing or not containing sperm whale echolocation click trains, derived from the original annotations. The 4-minute length was chosen as it was the finest common resolution across all sources. Additional anthropogenic and environmental information, such as the presence of other sounds like ships and other vocalising animals, was retained when available, although these annotations were solely kept as references to ensure a diverse range of anthropogenic and environmental noise sources in the final dataset but were not used as labels for training. 

Imprecisely labelled files were retained as examples of sperm whale absence only when species identification could definitively exclude sperm whales (e.g., files labelled as containing delphinids without any further species information).  In contrast, files with annotations that did not definitely rule out sperm whales (e.g., files labelled as containing unknown odontocete click trains) were excluded due to ambiguity.  Files labelled as containing single sperm whale clicks but not click trains were also removed, due to variability in this category between expert analysts \citep{garrobefonollosaComparingNeuralNetworks2024}. The final dataset comprised all available files containing sperm whale echolocation click trains, along with a randomly chosen set of files without them, ensuring a 50/50 split between presence and absence files.We chose to enforce a balanced split because the natural density of recordings with sperm whale click presence varies substantially between sources. Retaining original proportions would have made it difficult to obtain consistent comparisons across sources, so a uniform balance provided a clear benchmark. Absence files included a mix of files labelled as containing other impulsive noise sources, as well as some with no annotated instances.


In addition to the dataset of 4-minute segments, a dataset comprising 30-second sound files was constructed, using data from the AS and MED sources where annotations were available at a single click level (Table \ref{tab:sources}). . This 30-second dataset, when paired with the 4-minute one, allowed us to examine the trade-off between longer segments that provide greater temporal context and shorter segments that enable higher resolution feature extraction.

After selecting the 4-minute and 30-second segments and standardising the labels to presence/absence of sperm whale echolocation click trains for the whole segment, a pre-processing stage was implemented to standardise the recordings across diverse deployments. For multi-channel recordings, only the data recorded  furthest from the boat was retained. Furthermore, any DC offset present in the signal was removed. Subsequently, all files were decimated (with antialiasing) to a uniform sampling rate of 48 kHz . To eliminate low-frequency noise, the signal was filtered via a 1-20kHz bandpass 6-pole Butterworth filter, leaving a frequency range that captures the majority of the sperm whale click energy \citep{goold_time_1995}.

\subsection{Developing feature extraction methods} \label{subsec:featext}

Features were generated in two ways: (i) expert-based or handpicked feature selection and (ii) feature extraction via variational autoencoders (VAEs). Whilst the former is more transparent and rooted in marine science \citep{cox_development_2011}, it may not necessarily be optimised for the classifier. VAEs, in contrast, produce probabilistic latent representations of PAM segments, which are less intuitive to humans, but capable of capturing key temporal dynamics and structure in click trains. However, both of these methods operate on really short segments, as we wanted to capture local information. However, both methods operate on short segments, as we aimed to capture local information. Thus, a necessary first step was extracting local acoustic clips for feature engineering.

\subsubsection{Extraction of short clips for training feature extraction methods}
\begin{figure}[h!]
    \centering
    \includegraphics[width=0.9\textwidth]{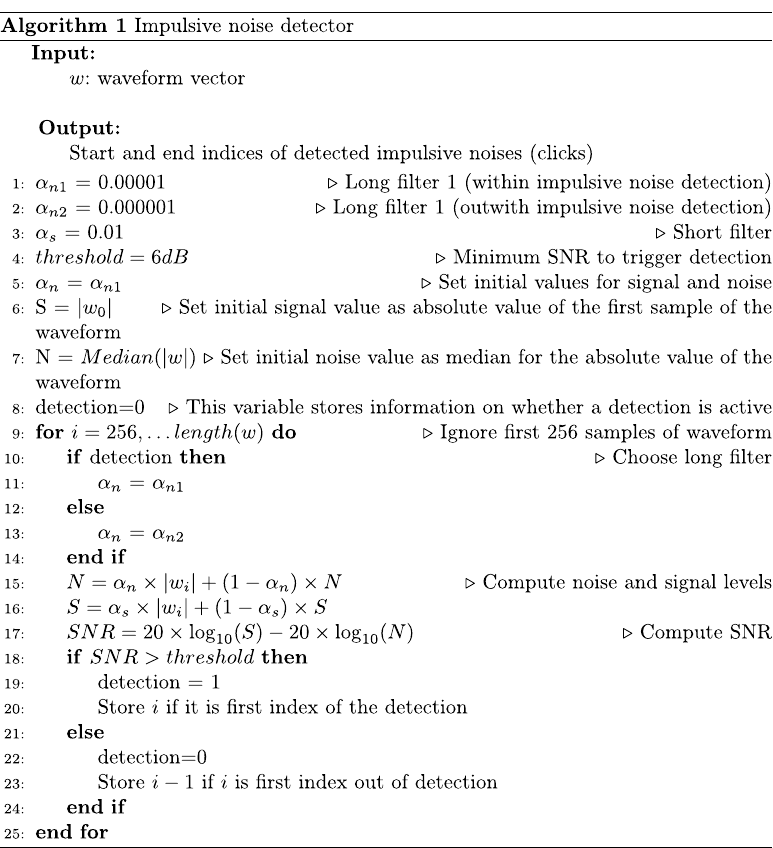}
   
\label{alg:impulsive}
\caption{Pseudocode for the impulsive noise detector used to identify prominent impulsive sounds from 4-minute recordings for variational autoencoder (VAE) training. The first 256 samples are discarded to ensure any detected clicks could be fully centred within an extraction window. }
\end{figure}

An impulsive noise detector algorithm was used to identify and select up to 10 impulsive sounds with the highest amplitude within each 4-minute segment, regardless of what the segment had been labelled as. Impulsive sounds were detected using a single-pole filter with a 6dB threshold   (see Alg. 1 shown in Figure~\ref{alg:impulsive}).  Subsequently,  waveform segments of varying lengths were extracted from these transients to train different feature extraction methods, ensuring the random placement of amplitude peaks within each segment. While impulsive noises detected in a 4-minute sequence may not necessarily share the label of the parent sound file, a considerable proportion are expected to do so, such that this procedure produces a dataset representative of the different transient sound sources present in the data. Additionally, 5 randomly sampled segments per file were included, to incorporate examples of background noise.  Random cropping of short segments within a 4-minute recording would rarely capture enough clicks and transient sounds for the model to be able to discriminate between them, so the strategy of combining impulsive noise detections and random segments therefore increases the likelihood of including sperm whale clicks and other confounding sources of transient noise, as well as representative background noise samples. 
In summary, for each 4-minute labelled segment, we extracted up to 10 high amplitude sequences plus 5 randomly selected quiet periods to train the feature extractors.

\FloatBarrier    
\subsubsection{Handpicked features}
\begin{figure}
    \centering
    \includegraphics[width=\textwidth]{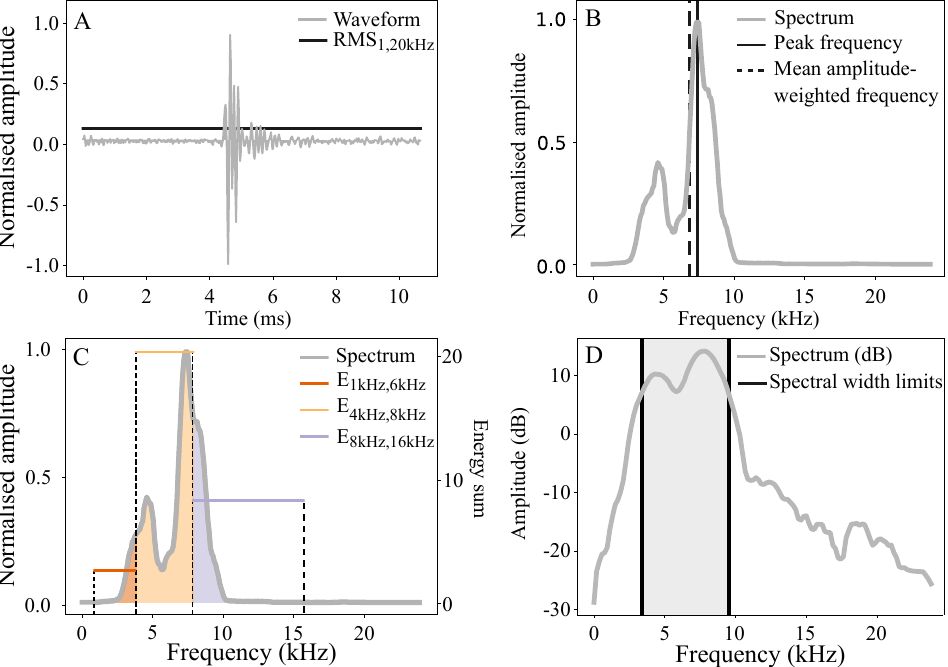}
    \caption{Parametrisation of a sperm whale (\textit{Physeter macrocephalus}) click using handpicked features (Table \ref{tab:handcrafted}), showing (A) the root-mean-square (RMS) over a fixed window, (B) mean amplitude-weighted and peak frequencies, (C) energy sum over different frequency bands, and (D) specral width of the signal, allowing an 8dB drop from the peak frequnecy at each side.  }
    \label{fig:click}
\end{figure}

Several acoustic features were selected from the non-overlapping short sound segments generated as described in Subsection~\ref{subsec:featext}, namely the root-mean-square of the time-series for a specific frequency band $[f_1, f_2]$ ($RMS_{f_1,f_2}$), mean amplitude-weighted and peak frequencies, spectral width, and energy sum over three frequency bands (1kHz - 4kHz, 4kHz-8kHz, 8kHz-16kHz; see Figure~\ref{fig:click} for an example and Table \ref{tab:handcrafted} for formal definition of each handpicked feature). The choice of these parameters was motivated by previous work on detection-classification of odontocete vocalisations  \citep{cox_development_2011}, and the frequency bands were chosen to correspond to the spectral profile of sperm whale clicks  \citep{goold_time_1995}. 

The handpicked features were combined into different sets of parameters (Table \ref{tab:combinations}). Combinations include: RMS only vectors over 1, 3, and 5 frequency bands, which were chosen for their ability to capture amplitude peaks across different frequency ranges; a set including 1 RMS band, the spectral features and the energy values, selected to incorporate amplitude, energetic and spectral information; and a subset of the previous one including only one RMS value along with peak frequency, mean amplitude-weighted frequency, and spectral width, which we chose to test whether dividing the signal into energy bandwidths was being used or simply providing redundant information.

\FloatBarrier
\subsubsection{Training Variational autoencoders (VAEs)}

VAEs were trained on three types of input data: waveforms, spectral profiles, and spectrogram representations. All VAEs were trained on the dataset of short clips described above, which includes both examples of different impulsive sounds and segments of background noise, ensuring the model trains on data representative of the acoustic variability relevant for sperm whale click detection tasks. To generate spectrograms and spectral profiles we used a 512-point Hann window FFT with a 50\% overlap. Spectral profiles were computed on segments of 512 samples, and frequencies below 1kHz or above 20kHz were discarded, leading to a final sequence of 200 amplitude estimations, which was then smoothed using an 8-point moving average filter. Additionally, longer spectrograms were computed for segments of 32,768 samples (0.68 seconds), yielding a spectrogram consisting of 128 FTTs sequences. Again, frequencies below 1kHz and above 20kHz were discarded, and the final representation was resized to 128$\times$128 pixels. 

The 2D-VAE trained to extract features from spectrograms used a residual network (ResNet-18) architecture \citep{stastny_vae-resnet18_2019}, which has consistently exhibited good performance in bioacoustic signal detection \citep{bergler_deep_2019, kirsebom_performance_2020, li_automated_2021, ryazanov_deep_2021}. Waveforms and smoothed spectral profiles were autoencoded using a 1D-VAE \citep{loris_1d_2019}. Each model was trained to compress data to different dimensions (latent sizes), specifically 24, 32, 48 and 64, to assess the impact of compression on feature quality. The 2D-VAE was implemented for two additional compression sizes, 96 and 128. Since VAEs return the parameters of a normal distribution in the latent space, for each input data point encoded into a space of dimension $n$, a vector of size 2$n$ is outputted where the first half of the vector corresponds to the means of the normal distribution and the second half to the variances. This strategy of mapping data to normal distribution parameters facilitates interpolation between data points, enhancing the robustness of embeddings (compressed representations of the VAEs) against various forms of noise and domain shifts \citep{rezende_stochastic_2014}. 

\FloatBarrier
\subsection{Training TCNs for the classification of sequences of extracted features}
For sound classification at the file level, recordings were divided into shorter, non-overlapping windows of two sizes: 512 and 2048 samples (0.01 s and 0.04 s, respectively). The  window lengths were used to investigate the trade-off between temporal resolution and sequence length. This process generated 22,500 and 5,625 windows, respectively, for 4-minute files, and 2,812 and 703 windows for 30-second files (Table \ref{tab:all_models}). The feature vectors detailed in the previous section were concatenated, transforming each sound file into a sequence of size ($m$, $n$), where $m$ represents the number of extracted features and $n$ denotes the number of non-overlapping windows in the file. Each of the $m$ features, both from handpicked features and VAE embeddings, was normalised using median and standard deviations of the feature on the values obtained on the short sound segment dataset. 

\begin{figure}
    \centering
    \includegraphics[width=\textwidth]{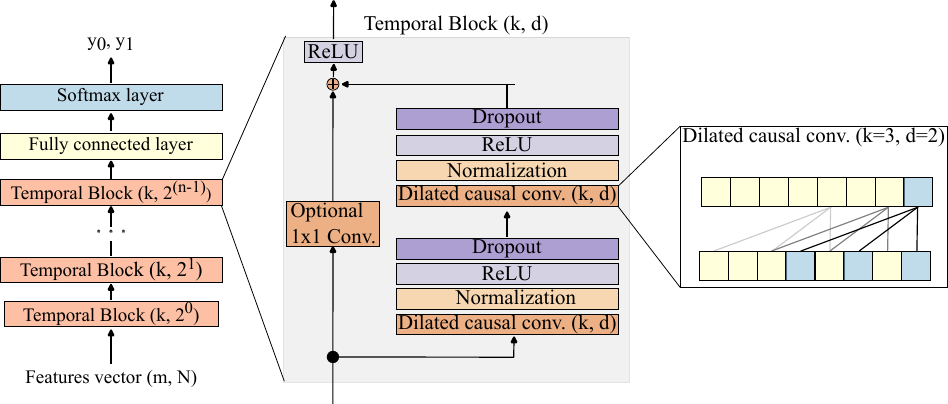}
    \caption{Diagram of the temporal convolutional network (TCN) architecture used to classify recordings based on the presence/absence of sperm whale (\textit{Physeter macrocephalus}) clicks (left), a temporal block (middle) and a dilated causal convolutional layer with a kernel size k = 3 and a dilation d = 2. The deep stack of dilated convolutions allows the network to capture long-range temporal patterns, making it a suitable architecture to detect the rhythmic vocalisation patterns of sperm whales. }
    \label{fig:tcn}
\end{figure}

A TCN was used for classifying full recordings from the sequence of extracted features based on the presence or absence of sperm whale clicks. The architecture shown in Figure \ref{fig:tcn} was adopted due to its proven ability to capture long-term dependencies and timing patterns in acoustic signals \citep{Lemaire2019TemporalCN}, which is important for detecting groups of sperm whales of varying sizes and distinguishing them from other sources of impulsive noise. The TCN was of length $n$ and had $m$ channels. 

All TCNs were trained in a supervised manner using the Adam optimiser \citep{kingma_adam_2017} and negative log likelihood loss function. Hyperparameters, including the number of hidden layers, kernel size, dropout rate, learning rate, and number of levels, were optimised using a grid search strategy (Table \ref{tab:hyperparameters}). The best-performing combination consisted of a batch size of 8, 8 hidden layers of size 10, a kernel size of 20, a dropout rate of 0.4, and a learning rate of 0.001.

\subsection{Comparison of trained TCN classifiers}
The performance of the trained TCNs were compared on an unseen subset of data reserved for testing. A random 70:15:15 split was used to partition the data into training, validation and testing sets. That split was consistent across models trained on the same length of recordings, and all three sets included data from all deployments. Performance evaluation was conducted based on recall ($TP/P$) and false-positive-rate ($FPR$; $FP/N$), where $P$ and $N$ represent the total number of positive (containing sperm whale clicks) and negative (not containing sperm whale clicks) files as annotated, and $TP$ and $FN$ denote correct and incorrect predictions of the model from the files labelled as containing sperm whales by humans. Model selection was guided by the trade-off between low $FPR$ and high recall.

\FloatBarrier
\section{Results}

\FloatBarrier
\subsection{Datasets composition}
The final 4-minute training set contained 7,668  sound files (Table \ref{tab:dataset}) originating from 8 sources, with a 50/50 split between files that were deemed as containing and not containing sperm whale clicks. This dataset included data from different sperm whale populations spanning 20 years, 8 geographical locations, 7 different recording devices and over 8 other vocalising species that produce echolocation clicks, as well as other sources of impulsive noise. The 30-second dataset contained  9,912 files from the two sources that provided click-level annotations, AS and MED (Table \ref{tab:dataset}).
\begin{table}[h!]
    \centering
    \caption{ Number of recordings in the sperm whale dataset by source and annotation category: sperm whales (SW) only, sperm whales with other impulsive noise, other impulsive noise only, and audited recordings with no annotations noted. Balearic recordings were only marked for sperm whale presence, so no information on other sound sources was therefore available. Numbers outside parentheses refer to the 4-minute dataset; numbers in parentheses refer to the 30-second dataset.}

    \begin{tabular}{cccccc}
\hline\
    Dataset & SW only & SW + Other & Other only & Nothing annotated & Total \\
    \hline
    AS & 149 (1,900) & 8  (26) & 123 (1,055)  & 234 (871) & 514  (3,852) \\   
    BAL\_1 & 93 & -& - &93  & 186 \\
    BAL\_2 & 143 & - & - &142 & 285  \\
    BAL\_3 & 273 & - & - & 273 & 546 \\
    CAL &173& 1 & 114 & 60 & 348 \\ 
    CS & 773 &560& 720 &607 & 2,660\\
    ICE & 365 & 558 & 379 & 544 & 1,846\\
    MED & 626 (3,003)& 16 (31)  & 47 (230) & 594 (2,796)& 1,283 (6,060)\\ 
    \hline
    Total & 2,595 (4,903)&  1,143 (57)& 1,383 (1,285)& 2,547 (3,667)& 7,668 (9,912)\\
    \hline
    \end{tabular}
    \label{tab:dataset}
\end{table}


A total of 101,210 short segments were extracted to train the feature extraction methods. 71,893 of these were sampled using an impulsive noise detector. Out of these, 37,638 came from recordings labelled as containing sperm whale clicks and 34,206 came from recordings labelled as not containing sperm whale clicks. The remaining 29,317 short sound segments were randomly sampled from the long sound segments to incorporate a representation of background noise. It is important to note that these latter segments were sampled independently from the impulsive noise detections. 

\FloatBarrier
\subsection{Feature extraction methods and training of 4-min audio file classifier}

\begin{figure}[h!]
    \centering
    \includegraphics[width=0.8\textwidth]{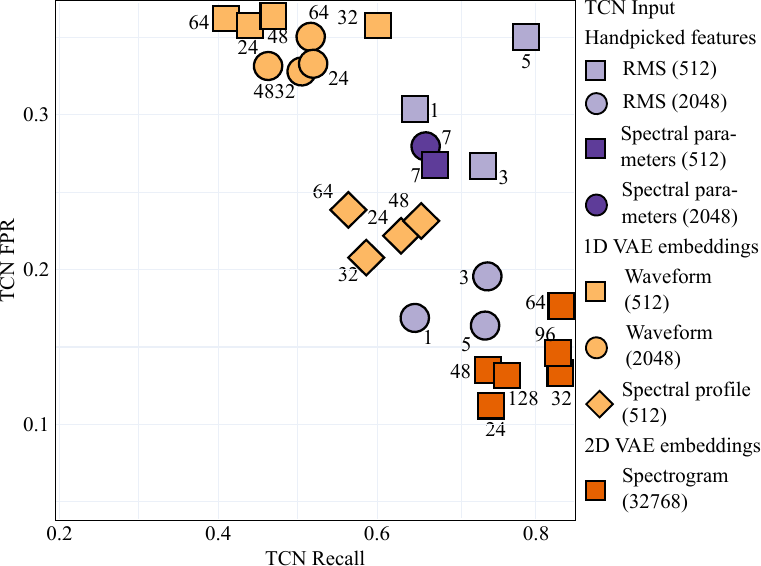}
    \caption{Recall vs false positive rate (FPR) on the 4-minute test set for temporal convolutional networks (TCNs) trained on variational autoencoder (VAE) embeddings and handpicked acoustic features. Each point represents a model configuration; the number of features per window is shown next to each point, and parentheses indicate the extraction window size. TCNs are used to classify recordings based on the presence/absence of sperm whale (\textit{Physeter macrocephalus}) click trains. The most efficient detectors are in the lower right quadrant, which in this case are the ones that work on sequences of VAE embeddings of spectrograms.}
    \label{fig:scatter}
\end{figure}

A total of 24 TCN models with optimised parameters and structure (Table \ref{tab:layers}) were trained on the collated 4-minute acoustic dataset for different forms of feature extraction (Figure \ref{fig:scatter}). The best performance in terms of high recall and low FPR was obtained using a 2D-VAE on spectrogram representations, followed by multi-channel RMS measures and a 1D-VAE on spectral profiles. TCNs applied to encoded representations of waveforms produced the poorest results every time, suggesting that the data entered into the VAE was too noisy and variable to extract the meaningful information needed for this task.

\begin{figure}[h!]
    \centering
    \includegraphics[width=.66\textwidth]{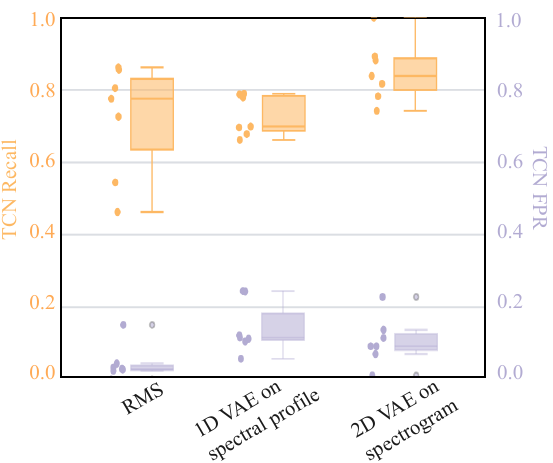}
    \caption{Box plot displaying recall (orange) and false positive rate (FPR, grey) scores on test data divided by deployment for the best-performing model across three feature extraction types. Dots represent each of the individual sources contributing to the sperm whale dataset (Table\ref{tab:sources}). Optimal models maximise recall while minimising FPR, resulting in clearly separated boxes. The figure shows that the detector using sequences of spectrogram embeddings from a variational autoencoder (VAE) performs best across deployments.  }
    \label{fig:boxes}
\end{figure}

Performance variability across sources shows that the TCN trained on 2D-VAE spectrogram data compressions also gave the highest minimum recall and the second lowest maximum FPR (Figure \ref{fig:boxes}). This suggests that this feature extraction method performs better overall and shows a more stable performance across data from different sources. 
\FloatBarrier

\subsection{The effect of annotation length on detector performance}
\begin{figure}[h!]
    \centering
    \includegraphics[width=0.8\textwidth]{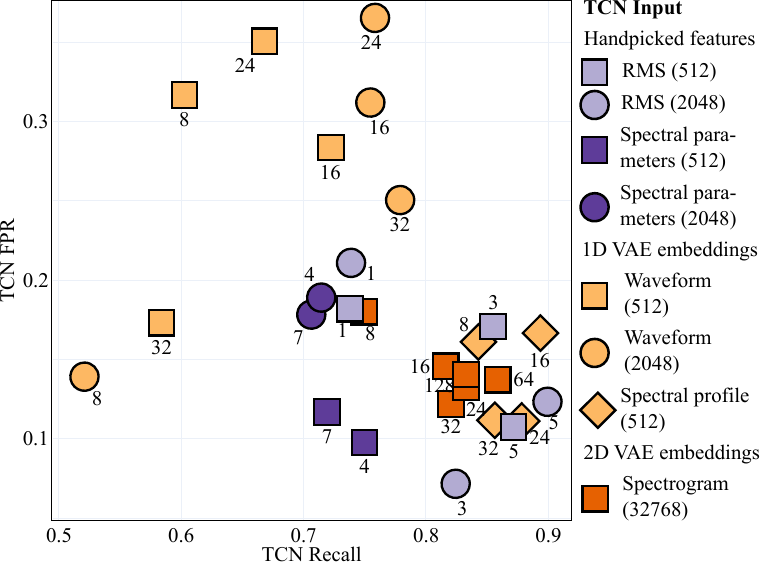}
    \caption{Recall vs false positive rate (FPR) on the 30-second test set for temporal convolutional networks (TCNs) trained on variational autoencoder (VAE) embeddings and handpicked acoustic features. Each point represents a model configuration; the number of features per window is shown next to each point, and parentheses indicate the extraction window size. TCNs are used to classify recordings based on the presence/absence of sperm whale (\textit{Physeter macrocephalus}) click trains. The most efficient detectors will be in the lower right quadrant, which in this case are the ones that work on sequences of RMS values over 5 frequency bands and the ones that work on sequences of VAE embeddings of spectral profiles.}
    \label{fig:30sec}
\end{figure}

\begin{figure}[h!]
    \centering
    \includegraphics[width=\textwidth]{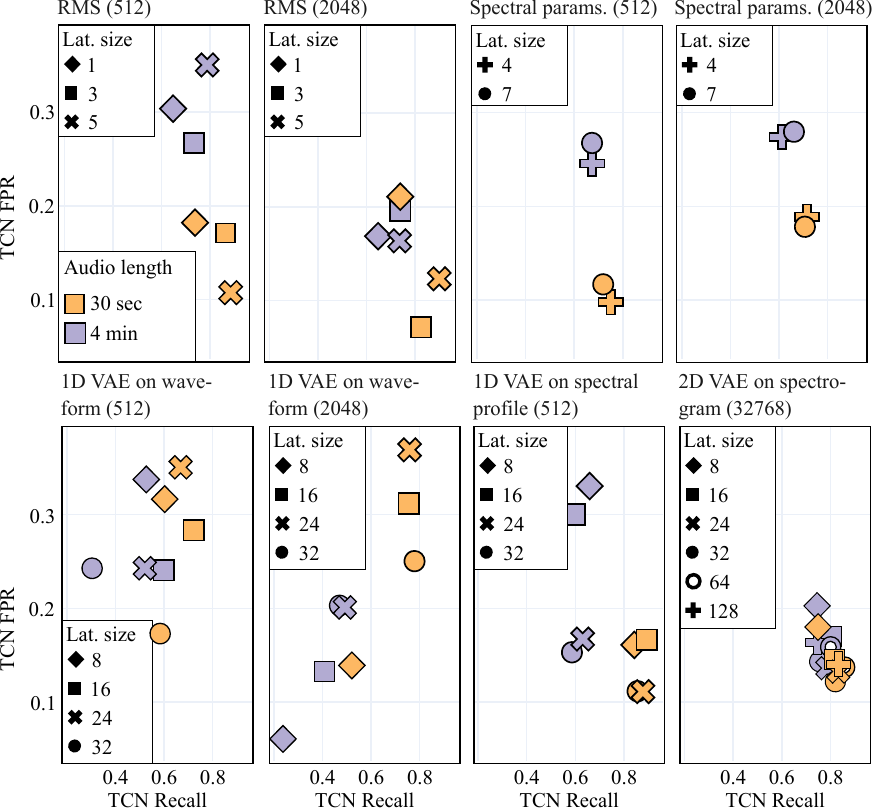}
    \caption{Recall versus false positive rate (FPR) on the validation set for temporal convolutional networks (TCNs) trained to detect sperm whale (\textit{Physeter macrocephalus})  on variational autoencoder (VAE) embeddings and handpicked acoustic features, shown for 30-second (orange) and 4-minute (grey) audio clips. Optimal detectors lie in the bottom right quadrant of each plot. TCNs generally perform better on shorter clips across most feature extraction methods. The exception is TCNs trained on VAE embeddings of spectrograms, which achieve consistent performance for both audio lengths.  }
    \label{fig:facet}
\end{figure}

For 30-second recordings, the results demonstrate a notable improvement in terms of higher recall and lower FPR for TCNs with different input features when compared to the features from the 4-minute recordings (Figure \ref{fig:30sec}, Figure \ref{fig:facet}). The only two exceptions to this trend are TCNs trained on VAE embeddings from waveforms, which exhibit no improvement and are consistently worse-performing method, and TCNs trained on VAE embeddings from spectrograms, which only show a marginal performance improvement. Notably, for 30-second recording classification, TCNs trained on VAE embeddings for spectral profile show the most improvement and achieve the best overall performance across all tested methods for feature extraction.

\FloatBarrier

\section{Discussion}
We developed a novel multi-step workflow to address the prevalent challenge of weakly labelled acoustic datasets in bioacoustic research. Our proposed approach consisted of four stages: (i) robust dataset curation from multiple geographical recordings to include various sources of environmental and anthropogenic noise, (ii) feature generation using expert-based feature selection and unsupervised feature extraction via VAEs, (iii) classification at the recording level using TCNs, which incorporates temporal context into the decision making. Our approach preserves the longer temporal dependencies inherent to bioacoustic data, in contrast to conventional strategies like the bag-of-frame method \citep{su_systematic_2014} or the strong label assumption \citep{hershey_cnn_2017}, both of which segment audio into short frames and discard broader temporal context. Our approach allowed TCNs to classify 4-minutes recordings at a performance level matching expert inter-annotator agreement rates \citep{garrobefonollosaComparingNeuralNetworks2024}, demonstrating that long-context, weakly labelled classification can approach the reliability of manual review.  

Our approach differs from previous methods for detecting sperm whale clicks by using TCNs to capture long-range dependencies at biologically-relevant timescales (e.g., minute-level click trains), moving beyond the isolated click-level analysis of traditional energy detectors \citep{kandia_detection_2006, morrissey_passive_2006, bermant_deep_2019}. By integrating VAEs as feature extractors, we achieve robust performance without manually set thresholds \citep{macaulay_passive_2020, webber_streamlining_2022} or finely labelled data \citep{garrobefonollosaComparingNeuralNetworks2024}, while also improving computational efficiency. This combination of long-range temporal modelling, reduced dependence on annotations, and efficient data compression addresses key limitations in current PAM pipelines.

We examined the effectiveness of handpicked acoustic features and VAEs for unsupervised feature extraction. Results showed that both methods provided sufficient information for a TCN to obtain recall rates of over 0.83 while maintaining a false detection rate below 0.13. TCNs trained on VAE embeddings of spectrogram representations performed marginally better on the task of classifying 4-minute recordings. Our results also showed that handpicked acoustic features were more prone to overfitting and exhibited greater variability in recall across sources, while VAE embeddings performed more consistently. This finding is particularly relevant to bioacoustic research as it highlights the capacity of VAEs to offer a consistent and reliable feature representation, regardless of variations in data sources and deployment conditions. 

To our knowledge, this study also presents the largest and most varied acoustic dataset of sperm whale clicks curated to date. It combines recordings from six sources, spanning decades, diverse geographic regions, and multiple deployment conditions, including recordings by multiple types of autonomous recorders as well as towed hydrophone arrays. This variability, which includes differences in background noise, vocalising species, and recording set-ups, is critical to training robust, generalisable models. 
The dataset was balanced to a 50/50 split of recordings containing and not containing sperm whale click trains within each source to allow for better comparison between sources. Real-world deployments, however, are often highly imbalanced, and future work applying this approach should account for such imbalances. We therefore chose FPR and recall as our primary evaluation metrics as, unlike precision or accuracy, they are inherently independent of class ratio and unaffected by the composition proportions of the dataset. They thus offer a more reliable and transferable measure of performance when applying the model to new datasets, provided one has a reasonable estimate of sperm whale density in the target environment. We hope that this dataset provides a reference point for future research on weakly-labelled bioacoustic detection, addressing a key need for open, high-quality bioacoustic baseline datasets \citep{frazao_workshop_2020}.

The benchmark nature of the dataset allowed us to explore differences in performance between data sources. We found that these could stem from variation in noise levels and recording conditions \citep{best_deep_2020, napoli_unsupervised_2023}, but annotation protocols and survey designs may also play an important role. For example, aurally auditing four minutes of single-channel data may lead to missed detections of fainter, less regular sperm whale clicks, whereas manually reviewing each click in isolation allows more thorough annotation. 

Moreover, the purpose of the study that generated a particular dataset also influences the detectability of the animals in the acoustic recordings. For example, the CS dataset was collected during a dedicated sperm whale behavioural study, where a vessel actively tracked whales to study social interactions. Consequently, the recordings predominantly contain high-amplitude, well-defined clicks from nearby individuals, introducing a proximity bias in the vocalisation examples. In contrast, datasets from static hydrophones or line-transect surveys are more likely to include faint, distant clicks, which poses challenges for both human annotators and automated detectors. When comparing detector performance across datasets, reported metrics reflect both the algorithm's efficacy and dataset-specific annotation biases. All results must therefore be interpreted within their recording and annotation context. 

This study focused on generalisability within data sources, as models were trained across a wide range of recording conditions and geographic regions, and evaluated on unseen recordings from those same sources, showing consistently strong performance. However, a limitation of the study is that we did not test cross-dataset transferability. This is an important consideration for deep learning applications in PAM, and the focus of a follow-up study \citep{garrobefonollosaMoreMerrierEffectsInpress}. 

In addition to generalisability, we also studied effect that segment duration had on performance, revealing that classifiers trained and tested on shorter-duration recordings (30 seconds instead of 4 minutes) consistently exhibited superior performance. Although this outcome was anticipated due to the reduced complexity of the task, the shorter dataset comprised recordings from only two sources, potentially contributing to lower data variation. However, for the 30-second dataset, TCNs trained on embeddings of spectral profiles outperformed those trained on the embeddings of 2D spectrograms, which were the best performing method for the 4-minute sound files. This suggests that the length of the final vector of acoustic features became a limiting factor in the classifier’s performance for 4-minute recordings. Consequently, methods that generate shorter sequences yielded better performance in spite of the lower time resolution. However, for classification at shorter time scales, these features extracted at a higher resolution considerably benefit the TCN step in the classification task.

\section{Conclusion}

Our study leverages existing annotated bioacoustic datasets to train automatic classification models, effectively overcoming previous limitations associated with weak labels. Given the variability in the dataset in terms of data sources, annotators, and labelling practices, the performance obtained by the model approaches an upper bound for a detector that does not overfit to the data. 
Notably, our ability to train the model on a temporal scale as large as 4 minutes means that marine scientists can readily apply this approach to many available annotated datasets without additional auditing, enabling the processing of large amounts of data efficiently. This capability opens doors for researchers to analyse higher volumes of marine acoustic data. 
Future research should focus on transferability of the best-performing models on distinct unseen new sources.

\begin{acknowledgments}

This work was funded by the Scottish Universities Partnership for Environmental Research (SUPER) Doctoral Training Partnership (DTP), between the Universities of St Andrews and Strathclyde. We would like to express our gratitude to Dr Kirsten Young, Thomas Webber,  Asociación TURSIOPS, Dr Felicia Vachon, Dr Hal Whitehead,  Dr Paul J. Wensveen, Michelle Dutro, Caroline Haas, the Marine Conservation Research/International Fund for Animal Welfare, and Greenpeace International for supplying the essential data for this research. Collection of the Iceland data set was made possible by the Rannís Icelandic Research Fund (number 207081). We would also like to thank Dr Volker Deecke and the  anonymous reviewers who made many helpful comments earliers draft of this manuscript.
\end{acknowledgments}

\section*{AUTHOR DECLARATIONS}
\subsection*{Conflicts of Interest}
The authors have no conflicts to disclose.

\section*{Data availability}
The acoustic datasets are available on request from the corresponding author, LFG. The data are not publicly available due to size restrictions.

Source code and trained models are openly available at github.com/laiagf/WeakDetector.

\newpage

\section*{Appendix}

\begin{table}[]
    \caption{Expert-selected acoustic features, with corresponding definitions and formulas, used to parametrise recordings as a compact sequence of feature vectors for deep learning applications.}

    \centering
    \begin{tabular}{p{0.2\textwidth}p{0.4\textwidth}  p{0.3\textwidth} }
    \hline
         Feature & Formula & Parameters \\
         \hline
         \hline
         Root mean square     &   $ RMS_{f_1, f_2} = \sqrt{\frac{1}{n} \displaystyle\sum_{i=1}^n x\left[f_1, f_2\right]_i^2}$ & $n$: number of samples in window.\\  
         & & 
         $x\left[f_1, f_2\right]$: waveform filtered between the frequencies of $f_1$ and $f_2$. \\
         \hline
         Peak frequency & $pkf =argmax\left(s(f)\right)$    &
         $f$: vector of frequencies returned by fft \\
         & & $s$: power spectrum smoothed using a 5-point moving average  \\
         \hline
         Amplitude-weighted mean frequency & $\overline{f} = \frac{s \cdot f }{\sum s }$ & $f$: vector of frequencies returned by fft. Frequency range 1kHz-20kHz. \\
         & & $s$: power spectrum smoothed using a 5-point moving average\\
        \hline
        Energy sum & $E_{f_1, f_2} =\displaystyle\sum_{f\in[f_1, f_2)} s[f]$ & $s[f]$: smoothed power spectrum at frequency $f$ \\ \hline
        Spectral width & $\omega = f_{M} - f_{m}$ & $f_M$ and $f_m$: maximum and minimum frequencies with an amplitude above the peak amplitude minus 8dB and with no more than 3 frequency values in a row below that threshold in between the peak and those frequencies. \\
        \hline

    \end{tabular}
    \label{tab:handcrafted}
\end{table}

\begin{table}[]
    \centering
        \caption{Combinations of handpicked acoustic features used to process recordings and train temporal convolutional networks (TCNs) for sperm whale (\textit{Physeter macrocephalus}) click train detection. }
    \begin{tabular}{p{0.1\textwidth}p{0.3\textwidth}p{0.5\textwidth}}
    \hline
         Type & Number of parameters & Parameters  \\
         \hline\hline
         \multirow{3}{.1\textwidth}{RMS only} & 1 & $RMS_{1-20kHz}$ \\
         \cmidrule(lr){2-3}
          & 3 &  $RMS_{1-6kHz}$,  $RMS_{6-12kHz}$,   $RMS_{12-20kHz}$ \\
          \cmidrule(lr){2-3}
         & 5 &  $RMS_{1-2kHz}$,  $RMS_{2-4kHz}$,   $RMS_{4-8kHz}$, $RMS_{8-16kHz}$,   $RMS_{16-20kHz}$, \\
         \hline
         Spectral features 
         & 7 & $RMS_{1-20kHz}$, $pkf$, $\overline{f}$, $\omega$, $E_{1-4kHz}$, $E_{4-8kHz}$, $E_{8-16kHz}$\\
         \hline
    \end{tabular}

    \label{tab:combinations}
\end{table}

\begin{table}[h!]
    \centering
        \caption{Layers and parameters of the temporal convolutional network (TCN) used to classify recordings based on the presence/absence of sperm whale (\textit{Physeter macrocephalus}) clicks. $N$ denotes is the number of non-overlapping windows per recording,  and $m$ denotes the number of features extracted for each window. }

    \begin{tabular}{ccc}
    \hline
        Layer &  Output shape & Param \# \\
        \hline
        \hline
         \multirow{3}*{Temporal Block 1 (k = 20, d = 1, dropout = 0.4)} & 
         \multirow{3}*{[-1, 25, N]} &   m*500 + 25 \\       
        &    &  12,525 \\
        & &  25*(m+1)\\
        \hline 
        
       \multirow{2}*{ Temporal Block 2 (k = 20, d = 2, dropout = 0.4)} & \multirow{3}*{[-1, 25, N]} & 12,525 \\
        & & 12,525 \\
        \hline
        
       \multirow{2}*{ Temporal Block 3 (k = 20, d = 4, dropout = 0.4)} &  \multirow{2}*{[-1, 25, N]} & 12,525 \\
        & & 12,525 \\
        \hline
        
       \multirow{2}*{ Temporal Block 4 (k = 20, d = 8, dropout = 0.4)} & \multirow{2}*{[-1, 25, N]} & 12,525 \\
        & & 12,525 \\
        \hline       

       \multirow{2}*{ Temporal Block 5 (k = 20, d = 16, dropout = 0.4)} &  \multirow{2}*{[-1, 25, N]} & 12,525 \\
        & & 12,525 \\
        \hline

       \multirow{2}*{ Temporal Block 6 (k = 20, d = 32, dropout = 0.4)} &  \multirow{2}*{[-1, 25, N]} & 12,525 \\
        & & 12,525 \\
        \hline

       \multirow{2}*{ Temporal Block 7 (k = 20, d = 64, dropout = 0.4)} &  \multirow{2}*{[-1, 25, N]} & 12,525 \\
        & & 12,525 \\
        \hline

       \multirow{2}*{ Temporal Block 8 (k = 20, d = 128, dropout = 0.4)} &  \multirow{2}*{[-1, 25, N]} & 12,525 \\
        & & 12,525 \\
        \hline
        Fully Connected Layer & [-1, 2] & 52 \\
        \hline
       
    \end{tabular}
    \label{tab:layers}
\end{table}

\begin{table}[h!]
    \centering
    \caption{Hyperparameters and tested values during temporal convolutional network (TCN) optimisation for classifying recordings based on the presence/absence of sperm whale (\textit{Physeter macrocephalus}) click  trains. Chosen configuration is in bold.}
    \begin{tabular}{cc}
    \hline
        \textbf{Parameter} & \textbf{Values } \\ 
        \hline
         Batch size & 4, \textbf{8}, 10 12, 16 \\
         Size of hidden layers & 8, \textbf{10}, 12, 15, 20, 25 \\
         Number of hidden layers & 4, \textbf{8}, 12, 16, 20 \\ 
         Kernel size & 10, 15, \textbf{20}, 25 \\
         Dropout & 0.1, 0.2, 0.3, \textbf{0.4} \\
         Learning rate & 0.1, 0.01, \textbf{0.001}, 0.0001 \\
         \hline
    \end{tabular}

    \label{tab:hyperparameters}
\end{table}
\begin{table}[]
    \caption{Summary of input data and sizes for all TCN models trained and tested.}

    \centering
    \begin{tabular}{p{0.2\textwidth}p{0.15\textwidth}p{0.2\textwidth}p{0.15\textwidth}p{0.15\textwidth}}
    \hline
     Feature extraction method & N. channels ($m$)  & Feature extraction window size & 4-min TCN length ($n$)  & 30-sec TCN length ($n$)  \\
     \hline
    \hline
    
    \multirow{2}{.2\textwidth}{RMS} & \multirow{2}{.15\textwidth}{1, 3, 5} &   512 samples & 22500 & 2812 \\
    \cmidrule(lr){3-5}
    & & 2048 samples & 5625 & 703 \\
    \cmidrule(lr){1-5}
    
    \multirow{2}{.2\textwidth}{Spectral parameters} & \multirow{2}{.15\textwidth}{7} &   512 samples & 22500 & 2812 \\
    \cmidrule(lr){3-5}
    & & 2048 samples & 5625 & 703 \\
    \cmidrule(lr){1-5}
    \cmidrule(lr){1-5}
    \multirow{2}{.2\textwidth}{1D-VAE on waveform} & \multirow{2}{.15\textwidth}{8, 16, 24, 32} &   512 samples & 22500 & 2812 \\
    \cmidrule(lr){3-5}
    & & 2048 samples & 5625 & 703 \\
    \cmidrule(lr){1-5}
    
    1D-VAE on spectral profile& 8, 16, 24, 32& 512 samples & 22500 & 2812 \\
    \cmidrule(lr){1-5}

    2D-VAE on spectrogram & 8, 16, 24, 32, 64, 128 & 32768 samples & 352 & 44 \\
    \hline
    \end{tabular}
    \label{tab:all_models}
\end{table}


\FloatBarrier

\end{document}